\documentclass[twoside]{amsart}

\usepackage[hmargin={3cm,3cm},vmargin={3cm,3cm},includehead]{geometry}
\usepackage{amsmath,amssymb}
\usepackage{amsfonts}
\usepackage[mathscr]{eucal}
\usepackage{dsfont}

\usepackage{epic}
\usepackage{eepic}

\newcommand{\ds}{\displaystyle}
\def\EXP{\textrm{{\large e}}}
\newcommand{\uop}{\mathbf{u}}

\newcommand{\wop}{\mathbf{w}}
\newcommand{\xop}{{\mathbf{A}}}
\newcommand{\yop}{{\mathbf{A}^\dagger}}

\newcommand{\alg}{\mathcal{A}}

\newcommand{\ii}{\mathsf{i}}
\newcommand{\pt}{\mathbf{n}}

\newcommand{\Nop}{\mathbf{N}}

\newcommand{\one}{\mathbf{e}_1}
\newcommand{\two}{\mathbf{e}_2}
\newcommand{\thr}{\mathbf{e}_3}
\newcommand{\er}{\mathbf{r}}

\newcommand{\tvec}[2]{\genfrac{(}{)}{0pt}{}{#1}{#2}}

\begin{document}

\vspace{2cm}

\title[]{Quantization of three-wave equations}%
\author{S. Sergeev}%
\address{Department of Theoretical Physics (Research School of Physical Sciences and Engineering) \& Mathematical Science
Institute, Canberra ACT 0200, Australia}
\email{sergey.sergeev@anu.edu.au}

%\thanks{This work was supported by the Australian Research Council}%


%-----------------------------------------------------
\begin{abstract}
The subject of this paper is the consecutive procedure of
discretization and quantization of two similar classical
integrable systems in three-dimensional space-time: the standard
three-wave equations and less known modified three-wave equations.
The quantized systems in discrete space-time may be understood as
the regularized integrable quantum field theories. Integrability
of the theories, and in particular the quantum tetrahedron
equations for vertex operators, follow from the quantum auxiliary
linear problems. Principal object of the lattice field theories is
the Heisenberg discrete time evolution operator constructed with
the help of vertex operators.
\end{abstract}

\maketitle

\section*{Introduction}

The tree-wave equations is the example of a completely integrable
classical system in $2+1$ dimensional space-time. These equations
apply to many physical systems. The aim of this paper is the
formulation of an integrable quantum field theory corresponding to
the three-wave equations.

Given a classical theory with action $S[\phi]$, there is the
universal prescription for the quantization. The amplitude between
a state $\phi_{in}(\er)$ at time $t_1$ and a state
$\phi_{out}(\er)$ at time $t_2$ is
\begin{equation}\label{feynman}
\langle\phi_{out}|\phi_{in}\rangle\;=\;\int_{
\genfrac{}{}{0pt}{}{\phi(\er,t_1)=\phi_{in}(\er)}{\phi(\er,t_2)=\phi_{out}(\er)}}
\mathscr{D}\phi\ \EXP^{\frac{\ii}{\hbar} S[\phi]}
\end{equation}
The perturbation-theory approach to the definition of
$\mathscr{D}\phi$ we reject from every beginning. The practical
way to define the measure in the Feynman integral is the
discretization. It is to be understood as
\begin{equation}\label{measure}
\mathscr{D}\phi \;=\; \lim \prod_{i,j} d\phi(\er_i,t_j)\;,
\end{equation}
where the limit symbol stands for infinitely dense discretization
of the space-time $(\er,t)$.

Thus, the first step toward the field theory is the discretization
of classical system. Only then we may look for a self-consistent
Heisenberg quantum mechanics on the lattice and regard it as the
regularized field theory.

A schematic outlook of milestones of our method is the following.
A classical theory is defined by the time dynamics for a field
$A$, $\ds \frac{d}{dt} A = f[A]$ (space-like degrees of freedom
are omitted for brevity). The dynamics is generated by a
Hamiltonian, $f[A]=\{H,A\}$, where $\{,\}$ are properly defined
Poisson brackets. Corresponding discrete time dynamics is the
evolution transformation $A(t+\Delta t)=F[A(t)]$. For brevity, we
choose the scale $\Delta t=1$. The key point is that making the
discretization, we have to take care of the Hamiltonian structure
of the dynamics. The discrete time evolution must be a canonical
transformation, it should preserve properly defined Poisson
algebra of observable fields on space-like lattice. The quantum
algebra of observables is a result of Dirac quantization of the
Poisson algebra. The quantum evolution map $A(t)\to A(t+1)$ must
be the an automorphism of the algebra of observables allowing one
to define the Heisenberg evolution operator,
\begin{equation}
A(t+1)\;=\; U\ A(t) \ U^{-1}\;.
\end{equation}
Such scheme is the algebraic realization of the discrete measure
(\ref{measure}). Operator $U$ acts in a representation space of
the algebra of observables, the amplitude (\ref{feynman}) in the
Heisenberg form is
\begin{equation}\label{xxx}
\langle\phi_{out}|U^T|\phi_{in}\rangle =
\int_{\genfrac{}{}{0pt}{}{\phi_0=\phi_{in}}{\phi_{T}=\phi_{out}}}
\ \underbrace{\prod_{t=1}^{T-1} d\phi_t}_{\ds\mathscr{D}\phi} \
\underbrace{\prod_{t=1}^T
\langle\phi_{t}|U|\phi_{t-1}\rangle}_{\ds\EXP^{\frac{\ii}{\hbar}
S[\phi]}}\;.
\end{equation}

The discretization of the space-time, construction of the quantum
algebra of observables and the basis-invariant definition of
Heisenberg evolution operator for the standard and modified
three-wave systems is the subject of this paper.

\section{Three-wave equations}

We commence with a short reminding of the classical three-wave
equations in three-dimensional space. In what follows, we use the
short notations for the indices,
\begin{equation}
(i,j,k)=\textrm{any permutation of } (1,2,3)\;.
\end{equation}

\subsection{Standard three-wave equations}

Linear problem for the standard three-wave equations is the set of
six differential relations
\begin{equation}\label{3w-lp}
\partial_i\psi_j=A_{ij}\psi_i\;,\quad i\neq j
\end{equation}
for three auxiliary fields $\psi_i$. Consistency of (\ref{3w-lp})
gives the equations for the six fields $A_{ij}$:
\begin{equation}\label{3w-eom}
\partial_i A_{jk} = A_{ji}A_{ik}\;.
\end{equation}
These are the equations of motion for tree-wave resonant system
(see e.g. \cite{ZakharovManakov,Kaup}), or the three-wave
equations for the shortness.

\subsection{Modified three-wave equations}

There is another type auxiliary linear problem \cite{ttt}, the
second order differential equations for the scalar auxiliary field
$\phi$:
\begin{equation}\label{m3w-lp}
(\partial_i\partial_j - A_{ij}\partial_j - A_{ji}\partial_i +
A_{ij}A_{ji})\phi=0\;,\quad i\neq j\;.
\end{equation}
Consistency of (\ref{m3w-lp}) gives similar equations for the
fields $A_{ij}$,
\begin{equation}\label{m3w-eom}
\partial_i A_{jk} = (A_{ij}-A_{ik})(A_{jk}-A_{ji})\;.
\end{equation}
We call them the modified three-wave equations.

\subsection{Hamiltonians}

It is convenient to use the single alphabetical indices instead of
the numerical pairs. In this paper we will use the following
convention for both systems:
\begin{equation}\label{alphanum}
\begin{array}{lll}
\ds A_{12}^{}=A_a^{}\;, & \ds \quad A_{13}^{}=A_b^{}\;, & \ds
\quad A_{23}=A_c^{}\;,\\
\ds A_{21}^{}=A_a^{*}\;, & \ds \quad A_{31}^{}=A_b^{*}\;, & \ds
\quad A_{32}=A_c^{*}\;,
\end{array}
\end{equation}
Note, our notations are not cyclic with respect to $(1,2,3)$.

Equations (\ref{3w-eom}) and (\ref{m3w-eom}) are extremum
conditions for the action
\begin{equation}\label{action}
\mathcal{S}=\int d^3x \left( A_a^*\partial_3^{} A_a^{} -
A_b^*\partial_2^{} A_b^{} + A_c^*\partial_1^{} A_c^{} -V\right)
\end{equation}
where
\begin{equation}
V\;=\;A_a^*A_b^{}A_c^*-A_a^{}A_b^*A_c^{}
\end{equation}
for the standard three-wave equations (\ref{3w-eom}), and
\begin{equation}
V\;=\; (A_a^{}-A_b^{})(A_a^*-A_c^{})(A_b^*-A_c^*)
\end{equation}
for the modified three-wave equations (\ref{m3w-eom}). For a
moment, we ignore the reality conditions, the star in all these
notations \emph{does not} mean the complex conjugation.

The time derivative $\partial_t$ and space derivatives
$\partial_x,\partial_y$ for action (\ref{action}) may be chosen by
\begin{equation}\label{space-time}
\partial_1=\partial_t-\partial_x\;,\quad
\partial_2=-\partial_t\;,\quad
\partial_3=\partial_t-\partial_y\;.
\end{equation}
Let $\er=(x,y)$ stands for the space-like vector. The standard
Lagrange transform relating Lagrangians and Hamiltonians gives
\begin{equation}\label{hamiltonian}
\mathcal{H}\;=\;\int d^2\er \left( A_a^*\partial_y^{} A_a^{} +
A_c^*\partial_x^{} A_c^{} + V\right)\;,
\end{equation}
so that the equations of motion (\ref{3w-eom}) and (\ref{m3w-eom})
are
\begin{equation}\label{H-eom}
\frac{d}{dt}A_a^{}(\er,t)\;=\;\biggl\{\mathcal{H},A_a^{}(\er,t)\biggr\}\;,\quad
\textrm{etc.}
\end{equation}
where the Poisson brackets are defined by
\begin{equation}\label{Poiss}
\biggl\{A_v^*(\er',t),A_v^{}(\er,t)\biggr\}\;=\;\delta(\er'-\er)\;,\quad
v=a,b,c\;.
\end{equation}
Any other same-time bracket is zero.

Relations (\ref{alphanum}) and (\ref{space-time}) are just one of
many possible conventions for the field notations and space-time
separation. All such conventions in the continuous case are
equivalent. The choice (\ref{alphanum}) and the signs in
(\ref{space-time}) at this moment may be considered as an odd
decision of the author.

\section{The Discretization}

Now we proceed to the discrete analogue of the standard and
modified three-wave equations (\ref{3w-eom},\ref{m3w-eom}). The
straightforward discretization of three-dimensional space gives
the cubic lattice
\begin{equation}\label{LATTICE}
\mathbf{x}=x_1\one + x_2\two + x_3 \thr\;,\;\;
x_i\in\mathbb{R}\quad \mapsto\quad \pt = n_1\one + n_2\two +
n_3\thr\;, \;\; n_i\in\mathbb{Z}
\end{equation}
The discrete analogue of derivative is the difference derivative,
\begin{equation}
\partial_i\;\mapsto\; \Delta_i\;,\quad \Delta_i \phi(\pt)  \;\stackrel{\textrm{def}}{=}\;
\phi(\pt+\mathbf{e}_i) - \phi(\pt)\;.
\end{equation}
The point is that we apply the straightforward discretization to
the linear problems (\ref{3w-lp},\ref{m3w-lp}), discrete equations
of motion must appear as the consistency conditions.

\subsection{Standard Three-wave equations}

The discrete linear problem corresponding to (\ref{3w-lp}),
\begin{equation}\label{3w-LP}
\Delta_i\psi_{j,\pt}\;=\; A_{ij}(\pt) \psi_{i,\pt}\;,
\end{equation}
provides the discrete equations of motion
\begin{equation}\label{3w-EOM}
A_{jk}(\pt+\mathbf{e}_i) \;=\; \frac{A_{jk}(\pt) + A_{ji}(\pt)
A_{ik}(\pt)}{1-A_{ji}(\pt)A_{ij}(\pt)}\;.
\end{equation}

\subsection{Modified Three-wave equations}

The discrete linear problem corresponding to (\ref{m3w-lp}),
\begin{equation}\label{m3w-LP}
\phi_{\pt+\mathbf{e}_i+\mathbf{e}_j}-
Q_{ji}(\pt)\phi_{\pt+\mathbf{e}_i} -
Q_{ij}(\pt)\phi_{\pt+\mathbf{e}_j} + Q_{ij}(\pt)
Q_{ji}(\pt)\phi_{\pt}=0\;,
\end{equation}
where
\begin{equation}
Q_{ij}(\pt)=1+A_{ij}(\pt)\;,\quad\textrm{etc.,}
\end{equation}
provides
\begin{equation}\label{m3w-EOM}
Q_{jk}(\pt+\mathbf{e}_i) \;=\;
\frac{Q_{ji}(\pt)Q_{ik}(\pt)+Q_{ij}(\pt)Q_{jk}(\pt)-Q_{ij}(\pt)Q_{ji}(\pt)}{Q_{ik}(\pt)}\;.
\end{equation}

\section{Constant time surface and evolution}

Discrete equations of motion (\ref{3w-EOM},\ref{m3w-EOM})
evidently define a sort of one-step evolution: the fields at point
$\pt+\mathbf{e}_i$ are expressed in the terms of the fields at
point $\pt$. However, an explicit form of a space-like discrete
surface and detailed definition of a discrete time corresponding
to our choice (\ref{space-time}) needs some discussion.

The key point is the geometrical structure of our discretization
and the geometrical structure of the linear problems
(\ref{3w-LP},\ref{m3w-LP}).

Let vector $\pt$ in (\ref{LATTICE}) stands for the vertex of the
cubic lattice. The cubic lattice consists of \emph{vertices}
$\pt$,  \emph{edges} $(\pt,\pt+\mathbf{e}_i)$, and \emph{faces}
$(\pt,\pt+\mathbf{e}_i,\pt+\mathbf{e}_j,\pt+\mathbf{e}_i+\mathbf{e}_j)$.
The linear variable $\psi_{i,\pt}$ of (\ref{3w-LP}) should be
associated with $(\pt,\pt+\mathbf{e}_i)$--edge, the linear
variable $\phi_\pt$ should be associated with $\pt$-vertex, and
the fields $A_{ij}(\pt),A_{ji}(\pt)$ should be associated with
$(\pt,\pt+\mathbf{e}_i,\pt+\mathbf{e}_j,\pt+\mathbf{e}_i+\mathbf{e}_j)$--face.
This is justified by the structure of the linear equations, see
Fig. \ref{fig-3w-square} and Fig. \ref{fig-m3w-square}. In what
follows, symbol $\alg_{v,\pt}$ stands for the collection of fields
on $(v,\pt)$-th face.
\begin{figure}[ht]
\begin{center}
\setlength{\unitlength}{0.25mm}
\begin{picture}(440,140)
\put(20,20){\begin{picture}(100,100) \Thicklines
\path(0,0)(100,0)(100,100)(0,100)(0,0)
 \put(0,100){\circle*{3}}
 \put(40,110){\scriptsize $\psi_{1,\pt}$}
 \put(30,-10){\scriptsize $\psi_{1,\pt+\two}$}
 \put(-30,45){\scriptsize $\psi_{2,\pt}$}
 \put(110,45){\scriptsize $\psi_{2,\pt+\one}$}
 \put(40,45){\scriptsize $\alg_{a,\pt}$}
 \put(-10,105){\scriptsize $\pt$}
 \put(105,105){\scriptsize $\pt+\one$}
 \put(-30,-10){\scriptsize $\pt+\two$}
\end{picture}}
\put(200,70){$\Leftrightarrow$} \put(250,70){$\ds
\begin{array}{l} \ds \psi_{1,\pt+\two} =
\psi_{1,\pt} + A_{21}(\pt) \psi_{2,\pt}\\
\ds \psi_{2,\pt+\one}=\psi_{2,\pt} + A_{12}(\pt) \psi_{1,\pt}
\end{array}$}
\end{picture}
\end{center}
\caption{Graphical representation of the elements of the linear
equations (\ref{3w-LP}): linear variables $\psi_{i,\pt}$ are
associated with the edges of cubic lattice, the fields
$\alg_{a,\pt}=\left(A_{12}(\pt),A_{21}(\pt)\right)$ are associated
with the face of cubic lattice.} \label{fig-3w-square}
\end{figure}
\begin{figure}[ht]
\begin{center}
\setlength{\unitlength}{0.25mm}
\begin{picture}(500,140)
\put(20,20){\begin{picture}(100,100) \Thicklines
\path(0,0)(100,0)(100,100)(0,100)(0,0)
 \put(0,100){\circle*{3}}
 \put(40,45){\scriptsize $\alg_{a,\pt}$}
 \put(-13,107){\scriptsize $\phi_{\pt}$}
 \put(105,105){\scriptsize $\phi_{\pt+\one}$}
 \put(-30,-10){\scriptsize $\phi_{\pt+\two}$}
 \put(105,-10){\scriptsize $\phi_{\pt+\one+\two}$}
\end{picture}}
\put(150,70){$\Leftrightarrow$} \put(180,70){$\ds
\phi_{n+\one+\two} -
Q_{21}\phi_{\pt+\one}-Q_{12}\phi_{\pt+\two}+Q_{12}Q_{21}\phi_{\pt}=0$}
\end{picture}
\end{center}
\caption{Graphical representation of the elements of the linear
equation (\ref{m3w-LP}): linear variables $\phi_{\pt}$ are
associated with the vertex of cubic lattice, the fields
$\alg_{a,\pt}=\left(Q_{12}(\pt),Q_{21}(\pt)\right)$ are associated
with the face of cubic lattice.} \label{fig-m3w-square}
\end{figure}

Take now up the auxiliary linear problem (\ref{3w-LP}) and more
detailed derivation of (\ref{3w-EOM}).  There are two ways to
express $\psi_{1,\pt+\two+\thr}$, $\psi_{2,\pt+\thr}$,
$\psi_{3,\pt+\one+\two}$ in the terms of $\psi_{1,\pt}$,
$\psi_{2,\pt+\one}$, $\psi_{3,\pt}$.  The one way is to use the
six relations
\begin{equation}\label{3w-LHS}
\begin{array}{l}
\Delta_3\psi_{1,\pt+\two}=A_{31}(\pt+\two)\psi_{3,\pt+\two},\\
\Delta_1\psi_{3,\pt+\two}=A_{13}(\pt+\two)\psi_{1,\pt+\two},
\end{array}\quad
\begin{array}{l}
\Delta_3\psi_{2,\pt}=A_{32}(\pt)\psi_{3,\pt},\\
\Delta_2\psi_{3,\pt}=A_{23}(\pt)\psi_{2,\pt},
\end{array}
\quad
\begin{array}{l}
\Delta_2\psi_{1,\pt}=A_{21}(\pt) \psi_{2,\pt},\\
\Delta_1\psi_{2,\pt}=A_{12}(\pt) \psi_{1,\pt}.
\end{array}
\end{equation}
The second way is to use the six other relations
\begin{equation}\label{3w-RHS}
\begin{array}{l}
\Delta_3\psi_{2,\pt+\one}=A_{32}(\pt+\one)\psi_{3,\pt+\one},\\
\Delta_2\psi_{3,\pt+\one}=A_{23}(\pt+\one)\psi_{2,\pt+\one},
\end{array}\quad
\begin{array}{l}
\Delta_2\psi_{1,\pt+\thr}=A_{21}(\pt+\thr)\psi_{2,\pt+\thr},\\
\Delta_1\psi_{2,\pt+\thr}=A_{12}(\pt+\thr)\psi_{1,\pt+\thr},
\end{array}
\quad
\begin{array}{l}
\Delta_3\psi_{1,\pt}=A_{31}(\pt) \psi_{3,\pt},\\
\Delta_1\psi_{3,\pt}=A_{13}(\pt) \psi_{1,\pt}.
\end{array}
\end{equation}
Graphical representation for (\ref{3w-LHS}) and (\ref{3w-RHS}) is
the left and right hand sides of Fig. \ref{fig-3w-hex}. The
collection of relations (\ref{3w-LHS}) and (\ref{3w-RHS})
correspond to the faces of the cube
$(\pt,\pt+\mathbf{e}_i,\pt+\mathbf{e}_i+\mathbf{e}_j,\pt+\one+\two+\thr)$
-- the discrete space-time consistency is the consistency around
the cube.
\begin{figure}[ht]
\begin{center}
\setlength{\unitlength}{0.25mm}
\begin{picture}(560,214)
\put(20,20){\begin{picture}(200,174) \Thicklines
% hexagon
\path(0,87)(50,174)(150,174)(200,87)(150,0)(50,0)(0,87)
% mersedes
\path(100,87)(50,174)\path(100,87)(50,0)\path(100,87)(200,87)
% coords
 \put(50,174){\circle*{3}}
 \put(100,87){\circle*{3}}
% points
 \put(40,180){\scriptsize $\pt$}
% \put(155,180){\scriptsize $\pt+\one$}
% \put(-35,85){\scriptsize $\pt+\thr$}
 \put(100,75){\scriptsize $\pt+\two$}
% outer edges
 \put(90,182){\scriptsize $\psi_{1}$}
 \put(90,-10){\scriptsize $\psi_{1,23}$}
 \put(0,130){\scriptsize $\psi_{3}$}
 \put(180,37){\scriptsize $\psi_{3,12}$}
 \put(180,130){\scriptsize $\psi_{2,1}$}
 \put(0,37){\scriptsize $\psi_{2,3}$}
% inner edges
 \put(125,95){\scriptsize $\psi_{1,2}$}
 \put(80,130){\scriptsize $\psi_{2}$}
 \put(80,37){\scriptsize $\psi_{3,2}$}
% variables
 \put(120,130){\scriptsize $\alg_{a,\pt}$}
 \put(120,44){\scriptsize $\alg_{b,\pt+\two}$}
 \put(40,82){\scriptsize $\alg_{c,\pt}$}
\end{picture}}
\put(320,20){\begin{picture}(200,174) \Thicklines
% hexagon
\path(0,87)(50,174)(150,174)(200,87)(150,0)(50,0)(0,87)
% mersedes
\path(100,87)(150,174)\path(100,87)(150,0)\path(100,87)(0,87)
% coords
 \put(50,174){\circle*{3}}
 \put(150,174){\circle*{3}}
 \put(0,87){\circle*{3}}
% points
\put(40,180){\scriptsize $\pt$} \put(155,180){\scriptsize
$\pt+\one$} \put(-35,85){\scriptsize $\pt+\thr$}
% outer edges
 \put(90,182){\scriptsize $\psi_{1}$}
 \put(80,-10){\scriptsize $\psi_{1,23}$}
 \put(0,130){\scriptsize $\psi_{3}$}
 \put(180,37){\scriptsize $\psi_{3,12}$}
 \put(180,130){\scriptsize $\psi_{2,1}$}
 \put(0,37){\scriptsize $\psi_{2,3}$}
% inner edges
 \put(123,120){\scriptsize $\psi_{3,1}$}
 \put(125,45){\scriptsize $\psi_{2,13}$}
 \put(40,77){\scriptsize $\psi_{1,3}$}
% variables
 \put(65,130){\scriptsize $\alg_{b,\pt}$}
 \put(60,44){\scriptsize $\alg_{a,\pt+\thr}$}
 \put(130,82){\scriptsize $\alg_{c,\pt+\one}$}
\end{picture}}
\end{picture}
\end{center}
\caption{Consistency around the cube for the linear problem of Fig
\ref{fig-3w-square}. Here $\psi_{1}=\psi_{1,\pt}$,
$\psi_{1,2}=\psi_{1,\pt+\two}$,
$\psi_{1,23}=\psi_{1,\pt+\two+\thr}$, etc. The left hand side
corresponds to Eqs. (\ref{3w-LHS}), the right hand side
corresponds to Eqs. (\ref{3w-RHS});
$\alg_{a,\pt}=\left(A_{12}(\pt),A_{21}(\pt)\right)$, etc.,
according to (\ref{alphanum}).} \label{fig-3w-hex}
\end{figure}

The way to introduce the discrete time is to identify the discrete
space-time fields in the left hand side of Fig. \ref{fig-3w-hex}
with the discrete time $t$, and identify the discrete space-time
fields in the right hand side of Fig. \ref{fig-3w-hex} with the
discrete time $t+1$. This exactly corresponds to
(\ref{space-time}):
\begin{equation}\label{d-space-time}
\one = \mathbf{e}_t-\mathbf{e}_x\;,\quad \two = -
\mathbf{e}_t\;,\quad \thr = \mathbf{e}_t-\mathbf{e}_y\;,
\end{equation}
and therefore
\begin{equation}\label{n-xyt}
\pt=t\mathbf{e}_t+x\mathbf{e}_x+y\mathbf{e}_y\;:\quad
x=-n_1,\;\;\; y=-n_3\;,\;\;\; t=n_1-n_2+n_3\;.
\end{equation}
The following table gives the correspondence between the initial
$A_{ij}(\pt)$ - notations and the space-time notations, convention
(\ref{alphanum}) is taken into account:
\begin{equation}\label{3w-in}
\begin{array}{ll}
\ds A_{12}^{}(\pt) = A_a^{}(x,y,t) = A_a^{}\;, & \ds
A_{21}^{}(\pt) = A_a^*(x,y,t) = A_a^*\;,\\
\ds A_{13}^{}(\pt+\two) = A_b^{}(x,y,t) = A_b^{}\;, & \ds
A_{31}^{}(\pt+\two) = A_b^{*}(x,y,t) = A_b^{*}\;,\\
\ds A_{23}^{}(\pt) = A_c^{}(x,y,t) = A_c^{}\;, & \ds
A_{32}^{}(\pt) = A_c^{*}(x,y,t) = A_c^{*}\;.
\end{array}
\end{equation}
Here the third columns are the shortened notations. The right hand
side of Fig. \ref{fig-3w-hex} implies thus
\begin{equation}\label{3w-out}
\begin{array}{ll}
\ds A_{12}^{}(\pt+\thr) = A_a^{}(x,y-1,t+1) = \overline{A}_a^{}\;,
& \ds
A_{21}^{}(\pt+\thr) = A_a^{*}(x,y-1,t+1) = \overline{A}_a^{\;*}\;,\\
\ds A_{13}^{}(\pt) = A_b^{}(x,y,t+1) = \overline{A}_b^{}\;, & \ds
A_{31}^{}(\pt) = A_b^{*}(x,y,t) = \overline{A}_b^{\;*}\;,\\
\ds A_{23}^{}(\pt+\one) = A_c^{}(x-1,y,t+1) = \overline{A}_c^{}\;,
& \ds A_{32}^{}(\pt+\one) = A_c^{*}(x-1,y,t+1) =
\overline{A}_c^{\;*}\;,
\end{array}
\end{equation}
Here the third columns are the shortened notations as well. The
consistency condition of (\ref{3w-LHS}) and (\ref{3w-RHS}),
completely equivalent to Eq. (\ref{3w-EOM}), may be rewritten in
shortened notations as
\begin{equation}\label{3w-themap}
\left\{\begin{array}{ll} \ds \overline{A}_a^{\;*} = A_a^* + A_b^*
A_c^{}\;, & \ds \overline{A}_a^{} = \frac{A_a^{}(1-A_c^{}A_c^*) +
A_b^{}A_c^*(1-A_a^{}A_a^*)}{1-\overline{A}_b^{}\overline{A}_b^{\;*}}\;,\\
\\
\ds \overline{A}_b^{\;*} = A_b^* (1-A_c^{}A_c^*) - A_a^*A_c^* \;,
&
\ds \overline{A}_b^{} = A_b^{}(1-A_a^{}A_a^*) - A_a^{}A_c^{}\;,\\
\\
\ds \overline{A}_c^{\;*} = \frac{A_c^*(1-A_a^{}A_a^*) +
A_a^{}A_b^*(1-A_c^{}A_c^*)}{1-\overline{A}_b^{}\overline{A}_c^{\;*}}\;,
& \ds \overline{A}_c^{} = A_c^{}+A_a^* A_b^{}\;.
\end{array}\right.
\end{equation}

In the absolutely similar way one may consider the discrete linear
problem of Fig. \ref{fig-m3w-square}. Corresponding picture is
Fig. \ref{fig-m3w-hex}.
\begin{figure}[ht]
\begin{center}
\setlength{\unitlength}{0.25mm}
\begin{picture}(560,214)
\put(20,20){\begin{picture}(200,174) \Thicklines
% hexagon
\path(0,87)(50,174)(150,174)(200,87)(150,0)(50,0)(0,87)
% mersedes
\path(100,87)(50,174)\path(100,87)(50,0)\path(100,87)(200,87)
% outer points
 \put(40,180){\scriptsize $\phi$}
 \put(-15,82){\scriptsize $\phi_3$}
 \put(40,-10){\scriptsize $\phi_{23}$}
 \put(150,180){\scriptsize $\phi_1$}
 \put(150,-10){\scriptsize $\phi_{123}$}
 \put(205,82){\scriptsize $\phi_{12}$}
% inner point
 \put(105,92){\scriptsize $\phi_2$}
% variables
 \put(120,130){\scriptsize $\alg_{a,\pt}$}
 \put(110,44){\scriptsize $\alg_{b,\pt+\two}$}
 \put(40,82){\scriptsize $\alg_{c,\pt}$}
\end{picture}}
\put(320,20){\begin{picture}(200,174) \Thicklines
% hexagon
\path(0,87)(50,174)(150,174)(200,87)(150,0)(50,0)(0,87)
% mersedes
\path(100,87)(150,174)\path(100,87)(150,0)\path(100,87)(0,87)
% outer points
 \put(40,180){\scriptsize $\phi$}
 \put(-15,82){\scriptsize $\phi_3$}
 \put(40,-10){\scriptsize $\phi_{23}$}
 \put(150,180){\scriptsize $\phi_1$}
 \put(150,-10){\scriptsize $\phi_{123}$}
 \put(205,82){\scriptsize $\phi_{12}$}
% inner point
 \put(83,92){\scriptsize $\phi_{13}$}
% variables
 \put(65,130){\scriptsize $\alg_{b,\pt}$}
 \put(60,44){\scriptsize $\alg_{a,\pt+\thr}$}
 \put(130,82){\scriptsize $\alg_{c,\pt+\one}$}
\end{picture}}
\end{picture}
\end{center}
\caption{Consistency around the cube for the linear problem of Fig
\ref{fig-m3w-square}. Here $\phi=\phi_\pt$,
$\phi_{1}=\psi_{\pt+\one}$, $\phi_{12}=\phi_{\pt+\one+\two}$, etc.
Convention (\ref{alphanum}) is taken into account,
$\alg_{a,\pt}=\left(Q_{12}(\pt),Q_{21}(\pt)\right)$, etc.}
\label{fig-m3w-hex}
\end{figure}

Introducing the same-time variables for the left hand side of Fig.
\ref{fig-m3w-hex}  analogously to (\ref{3w-in}),
\begin{equation}\label{m3w-in}
\begin{array}{ll}
\ds Q_{12}(\pt)=u_a(x,y,t)=u_a\;, & \ds
Q_{21}(\pt)=w_a(x,y,t)=w_a\;,\\
\ds Q_{13}(\pt+\two)=u_b(x,y,t)=u_b\;, & \ds
Q_{31}(\pt+\two)=w_b(x,y,t)=w_b\;,\\
\ds Q_{23}(\pt)=u_c(x,y,t)=u_c\;, & \ds
Q_{32}(\pt)=w_c(x,y,t)=w_c\;,
\end{array}
\end{equation}
and for right hand side of Fig. \ref{fig-m3w-hex} analogously to
(\ref{3w-out}),
\begin{equation}\label{m3w-out}
\begin{array}{ll}
\ds Q_{12}(\pt+\thr)=u_a(x,y-1,t+1)=\overline{u}_a\;, & \ds
Q_{21}(\pt+\thr)=w_a(x,y-1,t+1)=\overline{w}_a\;,\\
\ds Q_{13}(\pt)=u_b(x,y,t+1)=\overline{u}_b\;, & \ds
Q_{31}(\pt)=w_b(x,y,t+1)=\overline{w}_b\;,\\
\ds Q_{23}(\pt+\one)=u_c(x-1,y,t+1)=\overline{u}_c\;, &  \ds
Q_{32}(\pt+\one)=w_c(x-1,y,t+1)=\overline{w}_c\;,
\end{array}
\end{equation}
we get the consistency condition in the shortened notations:
\begin{equation}\label{m3w-themap}
\left\{\begin{array}{ll} \ds\overline{u}_a^{}=
\frac{u_aw_b+u_bw_c-u_bw_b}{w_c}\;,&\quad\quad
\ds\overline{w}_a^{}=\frac{w_aw_bu_c}{u_cw_c+w_aw_b-w_aw_c}\;,\\\\
\ds\overline{u}_b^{}=\frac{u_aw_a+u_bu_c-u_au_c}{w_a}\;,
&\quad\quad
\ds\overline{w}_b^{}=\frac{u_cw_c+w_aw_b-w_aw_c}{u_c}\;,\\\\
\ds\overline{u}_c^{}=\frac{w_au_bu_c}{u_aw_a+u_bu_c-u_au_c}\;,
&\quad\quad
\ds\overline{w}_c^{}=\frac{u_aw_b+u_bw_c-u_bw_b}{u_a}\;.
\end{array}\right.
\end{equation}
These relations are completely equivalent to Eq. (\ref{m3w-EOM}).

The derivation of the equations of motion as the consistency of
linear problem around the cube is similar to the consistency
approach to the integrable equations on quad-graphs \cite{suris}.
The consistency around the cube for two-dimensional quad-graph
equations is an analogue of the Yang-Baxter equation -- variables
in the left and right hand sides of two hexagons of Fig.
\ref{fig-3w-hex} are the same. The consistency in this paper
implies the different variables in the left and right hand sides,
it defines a map. Therefore it is an analogue of the \emph{local
Yang-Baxter} equation or the \emph{tetrahedral Zamolodchikov
algebra} \cite{LYBE,FTE,TZA}, see next section.

Equations (\ref{3w-themap}) and (\ref{m3w-themap}) express the
discrete space-time fields $\{\alg_{v}(x,y,t+1)\;:\;v=a,b,c;\;
x,y\in\mathbb{Z}\}$ in the terms of the fields
$\{\alg_{v}(x,y,t)\;:\;v=a,b,c;\; x,y\in\mathbb{Z}\}$. The
space-like discrete surface is the collection of hexagons of Fig.
\ref{fig-3w-hex} for all $x,y\in\mathbb{Z}$ and fixed $t$. This is
the \emph{honeycomb lattice}. Equations (\ref{3w-themap}) and
(\ref{m3w-themap}) therefore define the discrete time evolutions
of the fields situated at the faces of the honeycomb lattices.

Often in the literature the dual lattices are used. The fields are
associated to the edges of dual three-dimensional lattice and to
the vertices of its section -- dual two-dimensional lattice. The
lattice dual to the honeycomb one is called the \emph{kagome
lattice}.

\section{Linear problem as the zero curvature representation}

Equations (\ref{3w-themap}) and (\ref{m3w-themap}) are identically
equivalent to Eqs. (\ref{3w-EOM}) and (\ref{m3w-EOM})
correspondingly, but look more complicated since we reverse the
time direction of $\two$ in Eq. (\ref{d-space-time}). Our way to
introduce time and space-like coordinates
(\ref{space-time},\ref{d-space-time}) is not yet motivated.

Equations (\ref{3w-themap}) define the map of variables
$\alg_v^{}=(A_v^{},A_v^*)$ to
$\overline{\alg}_v^{}=(\overline{A}_v^{},\overline{A}_v^{\;*})$,
$v=a,b,c$. Analogously, equation (\ref{m3w-themap}) define the map
of variables $\alg_v^{}=(u_v^{},w_v^{})$ to
$\overline{\alg}_v^{}=(\overline{u}_v^{},\overline{w}_v^{})$,
$v=a,b,c$. Denote both these maps by the symbol
$\mathcal{R}_{abc}$. Formally, $\mathcal{R}_{abc}$ is the operator
acting in the space of functions of $\alg_a,\alg_b,\alg_c$:
\begin{equation}\label{map-def}
\forall\;\Phi=\Phi(\alg_a,\alg_b,\alg_c)\quad:\quad
\biggl(\mathcal{R}_{abc}\circ \Phi\biggr)(\alg_a,\alg_b,\alg_c)
\;\stackrel{\textrm{def}}{=}\;
\Phi(\overline{\alg}_a,\overline{\alg}_b,\overline{\alg}_c)\;.
\end{equation}
The point is that the maps (\ref{3w-themap}) and
(\ref{m3w-themap}) are two basic set-theoretical solutions of the
tetrahedron equation:
\begin{equation}\label{TE}
\mathcal{R}_{abc}\mathcal{R}_{ade}\mathcal{R}_{bdf}\mathcal{R}_{cef}
=
\mathcal{R}_{cef}\mathcal{R}_{bdf}\mathcal{R}_{ade}\mathcal{R}_{abc}\;.
\end{equation}
It may be verified straightforwardly, see the appendix.

Take up now the question: why the maps (\ref{3w-themap}) and
(\ref{m3w-themap}) satisfy the functional tetrahedron equation.
The reason is that the auxiliary linear problems defining the maps
of dynamical variables from the left hand sides of Figs.
\ref{fig-3w-hex} and \ref{fig-m3w-hex} to the right hand sides are
the \emph{correctly oriented} zero curvature representations of
three dimensional integrable model in discrete space-time.

Let me demonstrate this statement for the map (\ref{3w-themap}).
The linear equations for the single face, see Fig.
\ref{fig-3w-square},
\begin{equation}\label{psi12}
\psi_{1,\pt+\two}=\psi_{1,\pt} + A_{a}^{*} \psi_{2,\pt}\;,\quad
\psi_{2,\pt+\one}=\psi_{2,\pt} + A_{a}^{} \psi_{1,\pt}\;.
\end{equation}
may be rewritten in the matrix form as
\begin{equation}\label{X-def}
\tvec{\psi_{1,\pt}}{\psi_{2,\pt+\one}}\;=\;X[\alg_a] \;\cdot\;
\tvec{\psi_{1,\pt+\two}}{\psi_{2,\pt}}\;,\quad
X[\alg_a] \;=\; \left(\begin{array}{cc} 1 & -A_{a}^{*}\\
A_{a}^{} & 1-A_{a}^{}A_{a}^*\end{array}\right)\;.
\end{equation}
The linear equations for all three faces of the left hand side of
Fig. \ref{fig-3w-hex} may be written as
\begin{equation}
\tvec{\psi_1}{\psi_{2,1}}=X[\alg_a]\tvec{\psi_{1,2}}{\psi_1}\;,\quad
\tvec{\psi_{1,2}}{\psi_{3,12}}=X[\alg_b]\tvec{\psi_{1,23}}{\psi_{3,2}}\;,\quad
\tvec{\psi_2}{\psi_{3,2}}=X[\alg_c]\tvec{\psi_{2,3}}{\psi_{3}}
\end{equation}
with the same matrix function $X[\alg]$ (\ref{X-def}). Iterating
these matrix equations, we come to
\begin{equation}
\left(\begin{array}{c} \psi_1\\ \psi_{2,1}\\ \psi_{3,12}
\end{array}\right) = X_{12}[\alg_a] X_{13}[\alg_b] X_{23}[\alg_c]
\left(\begin{array}{c}
\psi_{1,23}\\\psi_{2,3}\\\psi_{3}\end{array} \right)\;,
\end{equation}
where $X_{ij}[\alg_v]$ are the three by three block-diagonal
matrices, $X_{ij}$ coincides with (\ref{X-def}) in $(ij)$-block
and has the unity in complimentary block. For instance,
\begin{equation}
X_{12}[\alg_a]=\left(\begin{array}{ccc} 1 & -A_a^* & 0 \\ A_a^{} &
1-A_a^{}A_a^* & 0 \\ 0 & 0 & 1\end{array}\right)\;.
\end{equation}
Analogously, the right hand side of Fig. \ref{fig-3w-hex} provides
\begin{equation}
\left(\begin{array}{c} \psi_1\\ \psi_{2,1}\\ \psi_{3,12}
\end{array}\right) = X_{23}[\overline{\alg}_a] X_{13}[\overline{\alg}_b] X_{12}[\overline{\alg}_c]
\left(\begin{array}{c}
\psi_{1,23}\\\psi_{2,3}\\\psi_{3}\end{array} \right)\;.
\end{equation}
Thus, the consistency of the linear problem around the cube is the
\emph{Korepanov} equation \cite{Korepanov,FTE}
\begin{equation}\label{KE}
X_{12}[\alg_a] X_{13}[\alg_b] X_{23}[\alg_c] \;=\;
X_{23}[\overline{\alg}_c] X_{13}[\overline{\alg}_b]
X_{12}[\overline{\alg}_a]\;.
\end{equation}
Using definition (\ref{map-def}), we may rewrite the Korepanov
equation in the form similar to tetrahedral Zamolodchikov algebra
\cite{TZA},
\begin{equation}\label{TA}
X_{12}[\alg_a] X_{13}[\alg_b] X_{23}[\alg_c] \;=\;
\mathcal{R}_{abc}\circ X_{23}[\alg_c] X_{13}[\alg_b]
X_{12}[\alg_a]\;.
\end{equation}
Tetrahedron equation (\ref{TE}) is the equivalence of
decompositions of the uniquely defined map
$\alg_v\to\overline{\overline{\alg}}_v$
\begin{equation}\label{quadrilateral}
X_{12}[\alg_a]X_{13}[\alg_b]X_{23}[\alg_c]X_{14}[\alg_d]X_{24}[\alg_e]X_{34}[\alg_f]=
X_{34}[\overline{\overline{A}}_f]
X_{24}[\overline{\overline{A}}_e]
X_{23}[\overline{\overline{A}}_c]
X_{14}[\overline{\overline{A}}_d]
X_{13}[\overline{\overline{A}}_b]
X_{12}[\overline{\overline{A}}_a]
\end{equation}
into two different sequences of elementary maps.

The main difference between (\ref{KE}) and the local Yang-Baxter
equation \cite{LYBE,FTE} is that the numerical indices in
(\ref{KE},\ref{TA}) correspond to the components of the
\emph{tensor sum} of one-dimensional vector spaces $\psi_i$. The
Korepanov equation comes from the linear problem directly,
therefore it is genuine multi-dimensional generalization of the
Lax representation.

There is no analogous matrix form for the zero curvature
representation of the modified three-wave equations. One has to
work directly with the sets of linear relations (\ref{m3w-LP}).
Nevertheless, the orientation of the faces in Fig.
\ref{fig-m3w-hex} is correct, and a set-of-linear-relations
analysis similar to Eq. (\ref{quadrilateral}) provides the
set-theoretical proof of the corresponding tetrahedron equation
\cite{S-symplectic,S-opus}.

\section{Poisson brackets for the fundamental maps}

The maps (\ref{3w-themap}) and (\ref{m3w-themap}) define the
discrete-time evolution on the honeycomb lattice. Evolution is the
Hamiltonian one if it preserves Poisson brackets.

One may verify it for the map (\ref{m3w-themap}), if
\begin{equation}\label{m3w-bra}
\biggl\{u_v,w_v\biggl\}=u_vw_v\;,\quad v=a,b,c
\end{equation}
and any other type bracket for $u_v,w_v$ is zero, then
\begin{equation}
\biggl\{\overline{u}_v,\overline{w}_v\biggr\}=\overline{u}_v\overline{w}_v\;,
\quad v=a,b,c
\end{equation}
and any other type bracket for $\overline{u}_v,\overline{w}_v$ is
zero. Therefore, (\ref{m3w-themap}) is the canonical map
\cite{S-symplectic}. Restoring the space structure according to
(\ref{m3w-in},\ref{m3w-out}), we come to the whole system of
same-time Poisson brackets conserved by the evolution:
\begin{equation}
\biggl\{ u_v(x,y,t),w_{v'}(x',y',t)\biggr\} \;=\;
\delta_{v,v'}\delta_{x,x'}\delta_{y,y'}u_v(x,y,t)w_v(x,y,t)\;,
\end{equation}
any other type same-time bracket is zero. Poisson algebra with
such structure of delta-symbols is called the \emph{ultra-local}
one.

The Poisson brackets for the evolution (\ref{3w-themap}) are not
ultra-local. To get the ultra-locality, we need to modify the
discrete linear problem (\ref{X-def}). Let there
\begin{equation}\label{X-def-correct}
X[\alg] \;=\; \left(\begin{array}{cc} K & -A^*\\ A &
K\end{array}\right)\;,\quad \alg=(A,A^*,K)\;,\quad K^2=1-AA^*\;.
\end{equation}
The modified map comes from the Korepanov equation (\ref{KE}):
\begin{equation}\label{3w-themap-correct}
\left\{\begin{array}{ll} \ds \overline{A}_a^{\;*} =
\overline{K}_b^{\;-1}(K_c^{}A_a^* + K_a^{}A_b^* A_c^{})\;, & \ds
\overline{A}_a^{} = \overline{K}_b^{\;-1} (K_c^{}A_a^{} +
K_a^{}A_b^{}A_c^*)\;,\\
\\
\ds \overline{A}_b^{\;*} = K_a^{}K_c^{}A_b^*  - A_a^*A_c^* \;, &
\ds \overline{A}_b^{} = K_a^{}K_c^{}A_b^{} - A_a^{}A_c^{}\;,\\
\\
\ds \overline{A}_c^{\;*} = \overline{K}_b^{\;-1}(K_aA_c^* +
K_c^{}A_a^{}A_b^*)\;, & \ds \overline{A}_c^{} =
\overline{K}_b^{\;-1}(K_a^{}A_c^{}+ K_c^{}A_a^* A_b^{})\;.
\end{array}\right.
\end{equation}
By definition (\ref{X-def-correct}),
$\overline{K}_v^2=1-\overline{A}_v^{}\overline{A}_v^{\;*}$.
Additional property of this map is
\begin{equation}
\overline{K}_a\overline{K}_b = K_a K_b\;,\quad
\overline{K}_b\overline{K}_c = K_b K_c\;.
\end{equation}
The map (\ref{3w-themap-correct}) satisfies the functional
tetrahedron equation and preserves the ultra-local Poisson
brackets \cite{ZTE}
\begin{equation}\label{osc-pois}
\biggl\{ A_v^{*},A_v^{}\biggr\} \;=\; K_a^2\;, \quad
\biggl\{K_v^{},A_v^{}\biggr\}=-\frac{1}{2} K_vA_v\;,\quad
\biggl\{A_v^{*},K_v^{}\biggr\}=-\frac{1}{2} K_v^{}A_v^*\;,
\end{equation}
any other type bracket is zero.

The principal advantage of the Poisson structure is that it allows
one to define the lattice actions with the help of functions
generating the canonical transformations. This subject is
technically complicated, we postpone it for future publications.

\section{Quantization}

The Poisson algebra (\ref{osc-pois}) is the quasi-classical limit
of $q$-oscillator algebra \cite{Kulish}
\begin{equation}\label{q-osc}
\xop\yop\;=\;1-q^{2\Nop+2}\;,\quad \yop\xop=1-q^{2\Nop}\;,\quad
\xop q^{\Nop} = q^{\Nop+1}\xop\;,\quad \yop
q^{\Nop}=q^{\Nop-1}\yop\;.
\end{equation}
The Poisson bracket is the limit of commutator when
$q^2=\EXP^{-\hbar}\to 1$, $\xop\to A$, $\yop\to A^*$ and
$q^\Nop\to K$.

The Poisson algebra $\{u,w\}=uw$ (\ref{m3w-bra}) is the
quasi-classical limit of the Weyl algebra \cite{BR}
\begin{equation}\label{weyl}
\uop\wop = q^2 \wop \uop
\end{equation}
with $\uop\to u$, $\wop\to w$ when $q\to 1$.

The whole algebra of observables, corresponding to the set of
classical fields $\alg_v(x,y)$ is thus the tensor power of the
local $q$-oscillator algebra $\alg=(\xop,\yop,q^\Nop)$ for the
quantized three-wave system; and the tensor power of the local
Weyl algebra $\alg=(\uop,\wop)$ for the modified three-wave
system. After the quantization, the indices $a,b,c$ of the
operators (or, more generally, the indices are $(v,x,y)$,
$v=a,b,c$, $x,y\in\mathbb{Z}$) stand for components of the tensor
power.

Quantum maps follow from the quantum linear problems. Quantized
version of (\ref{X-def}) is \cite{IJMMS}
\begin{equation}\label{3w-qlp}
\begin{array}{l}
|\psi_{1,\pt}\rangle \;=\; \lambda_a q^{\Nop_a}
|\psi_{1,\pt+\two}\rangle - \yop_a |\psi_{2,\pt}\rangle\;,\\
|\psi_{2,\pt+\one}\rangle \;=\; q^{-1}\lambda_a\mu_a
\xop_a|\psi_{1,\pt+\two}\rangle + \mu_a q^{\Nop_a} |\psi_{2,\pt}
\end{array}\ \Leftrightarrow\ X[\alg_a]\;=\;
\left(\begin{array}{cc} \lambda_a^{}q^{\Nop_a} & -\yop_a \\
q^{-1}\lambda_a\mu_a\xop_a & \mu_a q^{\Nop_a}
\end{array}\right)
\end{equation}
Here we replace the linear variables $\psi_i$ by vectors
$|\psi_i\rangle$ from a formal right module of the whole algebra
of observables. Parameters $\lambda_a,\mu_a$ are
$\mathbb{C}$-valued spectral parameters, we introduce them for the
sake of completeness. The consistency of linear problem of Fig.
\ref{fig-3w-hex} (equivalent to the quantum Korepanov equation
(\ref{KE})) may be solved with non-commutative coefficients, the
answer is \cite{ZTE,IJMMS}
\begin{equation}\label{3w-aut}
\left\{\begin{array}{ll}
 \ds \overline{\yop}_a = \frac{\lambda_c}{\lambda_b}
q^{-\overline{\Nop}_b} (q^{\Nop_c} \yop_a
+\frac{\lambda_a\mu_c}{q} q^{\Nop_a} \yop_b \xop_c)\;, & \ds
\overline{\xop}_a = \frac{\lambda_b}{\lambda_c}
q^{-\overline{\Nop}_b} (q^{\Nop_c} \xop_a +
\frac{q}{\lambda_a\mu_c}
q^{\Nop_a}\xop_b\yop_c)\;,\\
\\
\ds \overline{\yop}_b = \lambda_a\mu_c q^{\Nop_a+\Nop_c}\yop_b -
\yop_a\yop_c\;, &
\ds \overline{\xop}_b = \frac{q^2}{\lambda_a\mu_c} q^{\Nop_a+\Nop_c}\xop_b - \xop_a\xop_c\;,\\
\\
\ds \overline{\yop}_c = \frac{\mu_a}{\mu_b}
q^{-\overline{\Nop}_b}(q^{\Nop_a}\yop_c +\frac{\lambda_a\mu_c}{q}
q^{\Nop_c}\xop_a\yop_b)\;, & \ds \overline{\xop}_c =
\frac{\mu_b}{\mu_a} q^{-\overline{\Nop}_b}(q^{\Nop_a}\xop_c +
\frac{q}{\lambda_a\mu_c} q^{\Nop_c} \yop_a \xop_b)\;.
\end{array}\right.
\end{equation}
Here
$q^{2\overline{\Nop}_b}=1-\overline{\yop}_b\overline{\xop}_b$, in
addition
\begin{equation}
q^{\overline{\Nop}_a+\overline{\Nop}_b}=q^{\Nop_a+\Nop_b}\;,\quad
q^{\overline{\Nop}_b+\overline{\Nop}_c}=q^{\Nop_b+\Nop_c}\;,
\end{equation}
One may verify, the map (\ref{3w-aut}) is the \emph{automorphism}
of the tensor cube of $q$-oscillator algebra (\ref{q-osc}).

Quantized version of the linear equation (\ref{m3w-LP}) for the
modified three-wave system is \cite{S-opus}
\begin{equation}\label{m3w-qlp}
\varkappa_a^{} |\phi_{\pt+\one+\two}\rangle - q\wop_a
|\phi_{\pt+\one}\rangle - \uop_a |\phi_{\pt+\two}\rangle +
\uop_a\wop_a|\phi_{\pt}\rangle = 0\;.
\end{equation}
Here $\varkappa_a$ is a $\mathbb{C}$-valued spectral parameter.
Solution of the consistency condition, see Fig. \ref{fig-m3w-hex},
gives
\begin{equation}\label{m3w-aut}
\left\{\begin{array}{lll} \ds
\overline{\uop}_a^{}=\Lambda_2^{}\wop_c^{-1}\;, & \ds
\overline{\uop}_b^{} = \Lambda_1^{}\uop_c^{}\;, & \ds
\overline{\uop}_c^{} = \uop_b^{}\Lambda_1^{-1}\;,\\
\ds \overline{\wop}_a^{} = \wop_b^{}\Lambda_3^{-1}\;, & \ds
\overline{\wop}_b^{} = \Lambda_3^{}\wop_a^{}\;, & \ds
\overline{\wop}_c^{} = \Lambda_2^{}\uop_a^{-1}\;,
\end{array}\right.
\end{equation}
where
\begin{equation}
\begin{array}{l}
\ds \Lambda_1\;=\; \uop_a^{}\uop_c^{-1} - q \uop_a^{}\wop_a^{-1} +
\varkappa_a^{} \uop_b^{}\wop_a^{-1}\;,\\
\ds \Lambda_2 \;=\; \frac{\varkappa_a}{\varkappa_b}
\uop_b^{}\wop_c^{} + \frac{\varkappa_c}{\varkappa_b}
\uop_a^{}\wop_b^{} - q^{-1}
\frac{\varkappa_a\varkappa_c}{\varkappa_b} \uop_b^{}\wop_b^{}\;,\\
\ds \Lambda_3\;=\;
\wop_a^{-1}\wop_c^{}-q\uop_c^{-1}\wop_c^{}+\varkappa_c^{}
\uop_c^{-1}\wop_b^{}\;.
\end{array}
\end{equation}
One may verify, the map (\ref{m3w-aut}) is the automorphism of the
tensor cube of the Weyl algebra (\ref{weyl}).

Both automorphisms (\ref{3w-aut}) and (\ref{m3w-aut}) satisfy the
``functional'' tetrahedron equation. In irreducible
representations they are the internal automorphisms,
\begin{equation}\label{conjugation}
\mathcal{R}_{abc}\circ\Phi \;\equiv\; R_{abc}^{}\; \Phi\;
R_{abc}^{-1}\;.
\end{equation}
Corresponding operators $R_{abc}$ satisfy the quantum
(operator-valued) tetrahedron equations. Matrix elements of
$R_{abc}$ are functions of the spectral parameters
$\lambda_v,\mu_v$ for (\ref{3w-aut}) and $\varkappa_v$ for
(\ref{m3w-aut}).

The local maps (\ref{3w-aut}) and (\ref{m3w-aut}) define the
evolution map on the honeycomb lattice via the identification
\begin{equation}\label{q-ev-map}
\begin{array}{l} \alg_a=\alg_{a,x,y}(t) \\  \alg_b=\alg_{b,x,y}(t)
\\
\alg_c=\alg_{c,x,y}(t)\end{array} \quad  \to\quad
\begin{array}{l}
\overline{\alg}_a=\alg_{a,x,y-1}(t+1)\\
\overline{\alg}_b=\alg_{b,x,y}(t+1)\\
\overline{\alg}_c=\alg_{c,x-1,y}(t+1)
\end{array}
\end{equation}
in accordance with (\ref{3w-in},\ref{3w-out}) and
(\ref{m3w-in},\ref{m3w-out}). The evolution map is the
automorphism of the whole algebra of observables. Parameters
$\lambda_v,\mu_v$ for the $q$-oscillator model and $\kappa_v$ for
the Weyl algebra model should be $(x,y)$-independent. Then, in
proper representations, the evolution is the internal automorphism
given by an evolution operator,
\begin{equation}\label{U-operator}
\Phi(t+1)= U \Phi(t) U^{-1}\;.
\end{equation}
Matrix elements of $U$ may be constructed with the help of matrix
elements of local $R_{abc}$.

Examples of irreducible representations providing a ``good''
quantum mechanics are the following. For the $q$-oscillator
algebra it is the case of real $q^2=\EXP^{-\hbar}$, $0<q<1$, and
the Fock space: the Fock vacuum is defined by $\xop|0\rangle
=\Nop|0\rangle =0$. The dagger of $\yop$ stands for the Hermitian
conjugation, $\Nop^\dagger=\Nop$. If in addition
$\ds\frac{\lambda_c}{\lambda_b}$, $\ds \frac{\mu_a}{\mu_b}$ and
$\ds\frac{\lambda_a\mu_c}{q}$ in (\ref{3w-aut}) are unitary
parameters, then $R_{abc}$ and $U$ are the well defined unitary
operators. Matrix elements of $R_{abc}$ are given in \cite{ZTE}.
Evaluation (\ref{xxx}) of $U^T$ in the framework of normal symbols
gives the Feynman-type integral in the discrete space-time. In the
quasi-classical limit $\hbar\to 0$ ($q^2=\EXP^{-\hbar}$), it may
be shown
\begin{equation}
\langle A^*_{out}|U^T|A^{}_{in}\rangle \;=\; \int \mathscr{D}A
\mathscr{D}A^* \ \EXP^{\frac{\ii}{\hbar} S[A,A^*]}\;,
\end{equation}
where $S[A,A^*]$ is the lattice action mentioned in the previous
section, and the measure of integration is as well the lattice
one.

The proper quantum mechanical representation of the Weyl algebra
is given by the modular dualization \cite{Faddeev}. In addition to
the given local Weyl pairs
\begin{equation}
\uop=\EXP^{P}\;,\quad \wop = \EXP^{Q}\;,\quad [Q,P]=\ii\hbar \
\Rightarrow\ q^2=\EXP^{-\ii\hbar}\;,
\end{equation}
it is necessary to consider the dual local pairs
\begin{equation}
\uop'=\EXP^{\frac{2\pi}{\hbar} P}\;,\quad
\wop'=\EXP^{\frac{2\pi}{\hbar} Q}\;,\quad q^{\prime 2}=\EXP^{-\ii
\frac{(2\pi)^2}{\hbar}}\;.
\end{equation}
The sets of equations (\ref{m3w-aut}) and similar equations for
dual pairs define the kernel of $R_{abc}$ unambiguously.  It is
known, in this case $R_{abc}$ and $U$ are well defined unitary
operators. The $PQ$-symbol of $U^T$ in the quasi-classical limit
$\hbar\to 0$ is
\begin{equation}
\langle P_{out}|U^T|Q_{in}\rangle \;=\; \int \mathscr{D}P
\mathscr{D}Q \ \EXP^{\frac{\ii}{\hbar} S[P,Q]}
\end{equation}
where the measure and the action are the lattice ones.

\section{Conclusion}

The honeycomb lattice evolution map (\ref{q-ev-map}) defined with
the help of local automorphisms (\ref{3w-aut}) end
(\ref{m3w-aut}), and the proper choice of the Hilbert space define
unambiguously the evolution operator for the lattice approximation
of corresponding quantum field theory. These quantum field
theories are integrable since the underlying quantum auxiliary
linear problems (\ref{3w-qlp}) and (\ref{m3w-qlp}) provide the
existence of the complete set of the integrals of motion
\cite{Korepanov,S-transfer,S-linear,S-letter,IJMMS}.

The one-step evolution (\ref{q-ev-map}) is the discrete form of
the Hamiltonian flow (\ref{hamiltonian},\ref{H-eom}). The quantum
lattice Hamiltonian $\mathbf{H}$ is defined by
\begin{equation}\label{q-ham}
U\;=\;\EXP^{-\, \frac{\ii}{\hbar} \mathbf{H} \Delta t}\;,
\end{equation}
where $\Delta t=\Delta x=\Delta y$ is the lattice spacing
parameter. In the continuous $\Delta t\to 0$ classical $q\to 1$
limit the quantum lattice Hamiltonian becomes exactly
(\ref{hamiltonian}). However, on the lattice $\Delta t$ is finite
(in this paper we used the scale $\Delta t=1$), and therefore
$\mathbf{H}$ is not a polynomial in the algebra of observables.
The evolution operator is the principal object of the field theory
rather than a Hamiltonian.

The fundamental problem of the field theory is the calculation of
the spectrum of evolution operator. This is the open question in
three-dimensional models. Spectral equations for the evolution
operators are not known yet (except for the two-dimensional limit
of the three-wave system, \cite{S-laser}).

Let me conclude the paper with a discussion of the role of simplex
equations in quantum field theory. The linear problem is the
starting point of the integrability. The tetrahedron equation
(\ref{TE}) and the Yang-Baxter equation for the two-dimensional
models are the elementary consequences of linear problem as the
zero curvature representation. Nevertheless, the  simplex
configurations may be considered as fragments of the space-time
lattice, for instance
\begin{equation}\label{pf}
Z\;=\;\int \underbrace{d\phi_1'd\phi_2'
d\phi_3'}_{\ds\mathscr{D}\phi} \dots
\underbrace{\langle\phi_1^{}\phi_2^{}|R_{12}|\phi_1',\phi_2'\rangle
\langle\phi_1',\phi_3^{}|R_{13}|\phi_1'',\phi_3'\rangle
\langle\phi_2',\phi_3'|R_{23}|\phi_2'',\phi_3''\rangle}_{\ds
\EXP^{\frac{\ii}{\hbar} S[\phi]}} \dots
\end{equation}
Here for brevity we consider a two-dimensional theory and the
triangle configuration. Such partition function corresponds to the
Feynman integral (\ref{feynman}) with the discrete measure
definition (\ref{measure}) and some particular discretization. The
Yang-Baxter equation (and the $D$-simplex equations in general) is
the condition of $Z$-invariance: the partition function (\ref{pf})
does not depend on particular details of the discretization, it is
an invariant function of the boundary fields only. Thus the
natural role of simplex equations in the lattice field theory is
the conditions of self-consistent definition of the measure
(\ref{measure}).

\noindent{\textbf{Acknowledgements}} I would like to thank all the
participants of SIDE VII conference for fruitful discussions.

\appendix

\section{Verification of the functional tetrahedron equations}

The functional tetrahedron equation for the map (\ref{3w-themap})
may be verified with the help of Maple 9.5 routine:
\begin{verbatim}
restart;

R:=proc(a,b,c,var) # x stands for A, y stands for A^*
 local X,Y;
 X[2]:= x[b]*(1-x[a]*y[a])-x[a]*x[c];
 Y[2]:= y[b]*(1-x[c]*y[c])-y[a]*y[c];
 X[1]:=(x[a]*(1-x[c]*y[c])+x[b]*y[c]*(1-x[a]*y[a]))/(1-X[2]*Y[2]);
 Y[1]:=y[a]+y[b]*x[c]; X[3]:=x[c]+y[a]*x[b];
 Y[3]:=(y[c]*(1-x[a]*y[a])+x[a]*y[b]*(1-x[c]*y[c]))/(1-X[2]*Y[2]);
 simplify(subs([x[a]=X[1],y[a]=Y[1],x[b]=X[2],y[b]=Y[2],x[c]=X[3],y[c]=Y[3]],var))
end;

TE:=var->R(1,2,3,R(1,4,5,R(2,4,6,R(3,5,6,var))))
        -R(3,5,6,R(2,4,6,R(1,4,5,R(1,2,3,var))));

for k from 1 to 6 do TE(x[k]); TE(y[k]); od;
\end{verbatim}

Verification of the tetrahedron equation for the map
(\ref{m3w-themap}) is
\begin{verbatim}
restart;

R:=proc(a,b,c,var)
 local U,W;
 U[1]:=(u[a]*w[b]+u[b]*w[c]-u[b]*w[b])/w[c];
 W[1]:=w[a]*w[b]*u[c]/(u[c]*w[c]+w[a]*w[b]-w[a]*w[c]);
 U[2]:=(u[a]*w[a]+u[b]*u[c]-u[a]*u[c])/w[a];
 W[2]:=(u[c]*w[c]+w[a]*w[b]-w[a]*w[c])/u[c];
 U[3]:=w[a]*u[b]*u[c]/(u[a]*w[a]+u[b]*u[c]-u[a]*u[c]);
 W[3]:=(u[a]*w[b]+u[b]*w[c]-u[b]*w[b])/u[a];
simplify(subs([u[a]=U[1],w[a]=W[1],u[b]=U[2],w[b]=W[2],u[c]=U[3],w[c]=W[3]],var))
end;

TE:=var->R(1,2,3,R(1,4,5,R(2,4,6,R(3,5,6,var))))
        -R(3,5,6,R(2,4,6,R(1,4,5,R(1,2,3,var))));

for k from 1 to 6 do TE(u[k]); TE(w[k]); od;
\end{verbatim}
\end{document}